\documentclass[useAMS,usenatbib]{mn2e} 
\usepackage{latexsym}
\usepackage{times} \usepackage{graphicx}
\usepackage[german,english,american]{babel} \newcommand{\apj}{ApJ}
\newcommand{\apjl}{ApJL} 
 \newcommand{\mnras}{MNRAS}

\def\lesssim{\mathrel{\hbox{\rlap{\hbox{\lower4pt\hbox{$\sim$}}}\hbox{$<$}}}}
\def\gtrsim{\mathrel{\hbox{\rlap{\hbox{\lower4pt\hbox{$\sim$}}}\hbox{$>$}}}}

\input epsf
\def\plotone#1{\centering \leavevmode
\epsfxsize=\columnwidth \epsfbox{#1}}

\newcommand{\beq}{\begin{equation}}
\newcommand{\eeq}{\end{equation}}
\newcommand{\beqa}{\begin{eqnarray}}
\newcommand{\eeqa}{\end{eqnarray}}

\def\aj{AJ}                   
             
\def\apj{ApJ}                 
\def\apjl{ApJ}                
\def\apjs{ApJS}

\def\mnras{MNRAS}

\title[Ram pressure stripping and cold fronts]{Ram pressure stripping and
the formation of cold fronts}

\author[Heinz et al.]{S.~Heinz,$^{1}$ E.~Churazov,$^{1,2}$ W.~Forman,$^{3}$
C.~Jones,$^{3}$ and U.G. Briel,$^{4}$\\
$^1$ Max-Planck-Institut f\"ur Astrophysik, Karl-Schwarzschild-Strasse 1,
85741 Garching, Germany\\
$^2$ Space Research Institute (IKI), Profsoyuznaya 84/32, Moscow 117997,
Russia\\
$^3$ Harvard-Smithsonian Center for Astrophysics, 60 Garden St., Cambridge,
MA 02138, USA \\
$^4$ MPI f\"{u}r Extraterrestrische Physik, P.O. Box 1603, 85740 Garching,
Germany }

\pagerange{\pageref{firstpage}--\pageref{lastpage}}
\pubyear{2003}

\begin{document}
\maketitle

\label{firstpage}
\begin{abstract}
Chandra and XMM-Newton observations of many clusters reveal sharp
discontinuities in the surface brightness, which, unlike shocks, have lower
gas temperature on the X-ray brighter side of the discontinuity. For that
reason these features are called ``cold fronts''.  It is believed that some
cold fronts are formed when a subcluster merges with another cluster and
the ram pressure of gas flowing outside the subcluster gives the contact
discontinuity the characteristic curved shape.  While some edges may not
arise directly from mergers (e.g., A496, \citealt{dupke:03}), this paper
focuses on those which arise as contact discontinuities between a merging
subcluster and the ambient cluster gas.  We argue that the flow of gas past
the merging subcluster induces slow motions {\it inside} the cloud. These
motions transport gas from the central parts of the subcluster towards the
interface. Since in a typical cluster or group (even an isothermal one) the
entropy of the gas in the central regions is significantly lower than in
the outer regions, the transport of the low entropy gas towards the
interface and the associated adiabatic expansion makes the gas temperature
immediately inside the interface lower than in any other place in the
system, thus enhancing the temperature jump across the interface and making
the ``tip'' of the contact discontinuity cool. We illustrate this picture
with the XMM-Newton gas temperature map of the A3667 cluster.
\end{abstract}

\begin{keywords}
methods: numerical, galaxies: general, galaxies: clusters: individual:
A3667, X-rays: galaxies: clusters
\end{keywords}

\sloppypar

\section{Introduction}
Cold fronts were discovered as sharp features in the X-ray surface
brightness distribution in Chandra observations of the clusters A2142 and
A3667 (\citealt{markevitch:00,vikhlinin:01}, see also
\citealt{markevitch:02}). Similar features have now been found in several
other clusters \citep{sun:02,kempner:02}. A natural assumption that these
sharp features were due to shocks propagating through the gas was
immediately rejected since the measured gas temperature was lower on the
X-ray brighter side of the feature. It was suggested instead that the
observed edges were formed as a result of the flow of hotter gas around the
colder gas cloud, the front itself being the contact discontinuity
separating hot and cold gas \citep{markevitch:00,vikhlinin:01}. The motion
of the cloud with respect to the gas on larger scales can be understood,
for example, as the result of a merger of a subcluster with a more massive
and hotter cluster. And indeed, features resembling cold fronts were found
in numerical simulations of cluster formation \citep{nagai:03,bialek:02}.

In this short paper we argue that ablation of the gaseous cloud causes
characteristic differential motion of the gas {\it inside} the subcluster,
which transports the low entropy gas from the subcluster core towards the
contact discontinuity, thus enhancing the jump in temperature and surface
brightness across the discontinuity.

\section{Illustrative model}
\label{sec:simulations}
\subsection{Numerical Setup}
In order to study the global dynamics and temperature structure of cold
clouds, we ran a number of hydrodynamic simulations.  Initial conditions
were chosen to correspond roughly to those observed in A3667
\citep{vikhlinin:01}, however, we varied all relevant parameters within an
order of magnitude to confirm that the conclusions we draw from the
simulations are robust.

The simulations were performed using the FLASH hydro code
\citep{fryxell:00}. This code is an adaptive mesh refinement code using a
high precision PPM solver, explicit in time and formally accurate to 2nd
order.  To verify our results, we also ran a batch of simulations using the
ZEUS 3D code \citep{clarke:94,stone:92,stone:92b}. We restricted most of
the simulations to two dimensions, the bulk of the simulations assuming
axi-symmetry. We ran several test cases in a cartesian semi-infinite slab
geometry to confirm that the axi-symmetric assumption did not corrupt our
results.  Furthermore, we re-ran our fiducial simulation in 3D (at lower
resolution) to confirm that the results are not an artifact of enforced
symmetries.

We used an adiabatic equation of state.  Both codes used incorporate shock
capturing schemes.  However, to keep the simulations scale free, we did not
include the effects of radiative cooling.  For our fiducial simulation
presented in the following section, it turns out that cooling of the
coolest gas is at most marginally important over the simulation time span:
For our fiducial simulation (see below) the minimum cooling time anywhere
during the simulation is $t_{\rm cool,min} \sim 4\,{\rm Gyrs}$, compared to
a total simulation time of 4 Gyrs.  While this is clearly a simplification,
the effects of cooling will only increase the temperature contrast between
the low and high entropy regions and thus enhance the effects discussed in
this paper.

The initial setup consists of a stationary isothermal
atmosphere\footnote{We note that the assumption of isothermality is
idealistic - any realistic halo will presumably possess a stratified
atmosphere with low entropy gas at the center. However, for the purpose of
our study, isothermality is a conservative assumption, as it decreases the
temperature contrast that can be achieved and more clearly reveals the
nature of the dynamical effects we will be discussing.}  of cold, dense gas
and a wind of uniform hot, light gas moving at uniform speed relative to
the isothermal atmosphere.  The isothermal atmosphere is assumed to be in
hydrostatic equilibrium, following a beta model density profile of the form
\begin{equation}
	\rho_{r}=\rho_c\left[1 +
	\left(\frac{r}{r_c}\right)^2\right]^{-3\beta/2}
\end{equation}
where we chose $\beta=0.5$, $r_{\rm c}=250\,{\rm kpc}$, and $\rho_{\rm
c}=3.6\times 10^{-3}\,{\rm cm^{-3}}$.  For simplicity, we assume that the
gravitational field binding the cold gas is provided predominantly by dark
matter.  Thus, we keep the gravitational potential fixed throughout the
simulations.  The uniform gas is assumed to have an electron density
$\rho_x=4.6\times 10^{-4}\,{\rm cm^{-3}}$, with a relative velocity such
that its internal Mach number is $M=v_x/c_x=1.1$, while the Mach number
with respect to the sound speed of the cold cloud is $M_{\rm x}=1.6$.  We
used a temperature normalisation of 7.7 keV for the undisturbed external
medium \citep{vikhlinin:01} and 3.2 keV for the undisturbed isothermal
cloud, which sets the length and velocity scale of the simulation.

The computational volume was restricted to the range of $r \in
\left[0\,{\rm Mpc},2\,{\rm Mpc}\right]$ and $z \in \left[-2\,{\rm
Mpc},2\,{\rm Mpc}\right]$.  Increasing the physical dimensions of the
simulated volume by a factor of two did
not affect the  internal structure of the cloud
significantly.  The initial interface is located at the $z=-500\,{\rm kpc}$
surface.  Inflow boundary conditions were chosen for the $z=-2\,{\rm Mpc}$
boundary (forcing velocity, pressure, and density to $v_x$, $p_x$, and
$\rho_x$ respectively), outflow boundary conditions for the $z=2\,{\rm
Mpc}$ and the $r=2\,{\rm Mpc}$ boundaries, and reflective boundary
conditions along the $r=0\,{\rm Mpc}$ axis.

The effective (i.e., maximum) resolution within this box was varied between
$1024\times 512$ and $8192\times 4096$ without significant impact on the
global dynamics.  The bulk of the simulations were run at an effective
resolution of $2048\times 1024$ (corresponding to a cell size of
$1.95\,{\rm kpc}$), with a refinement depth of 8, while forcing the
resolution in the central $500\,{\rm kpc}$ to be at least $3.9\,{\rm
kpc}$. For a typical size of $\sim 500\,{\rm kpc}$ (as observed in A3667)
this implies that the cloud is safely resolved with more than 100
resolution elements across a cloud radius.

In order to evaluate the observed temperature distribution, we calculated
emission weighted temperature maps, using the observed background
temperature and background intensity around the cold front in A3667 (due to
the limited box size and assumed uniform external density we cannot
calculate the background emission from the host cluster
self-consistently). We assumed pure bremsstrahlung emission, which is a good
approximation above $T\gtrsim 2\,{\rm keV}$ and sufficient for our
purposes.

\subsection{Simulation results}
The simulations generally show the following temporal evolution: the
interface between the moving and the stationary gas drives a strong shock
into the stationary, dense gas, while a weak shock is driven into the hot
gas (depending on the initial velocity, this weak shock develops into a
compression wave or a weak bow shock upstream).  The interface between the
compressed/shocked gases is a contact discontinuity, which envelopes the
cold gas in roughly a bullet shape.

The front of this contact discontinuity is roughly spherical, while
Kelvin-Helmholtz instabilities and turbulence, induced by the tangential
velocity discontinuity between cold and hot gas, lead to the development of
large eddies \citep{mazzotta:02}.

The shock passage through the cold gas moves the material with the lowest
entropy and the highest density away from the center of the potential well.
After the shock passage, this leads to a reversal of the velocity pattern,
due to the gravitational field whose center is no longer co--spatial with
the lowest entropy, highest density material.  This field accelerates the
material against the direction of the background flow, transporting it back
towards the contact discontinuity.

\begin{figure*}
\resizebox{0.98\textwidth}{!}{\includegraphics{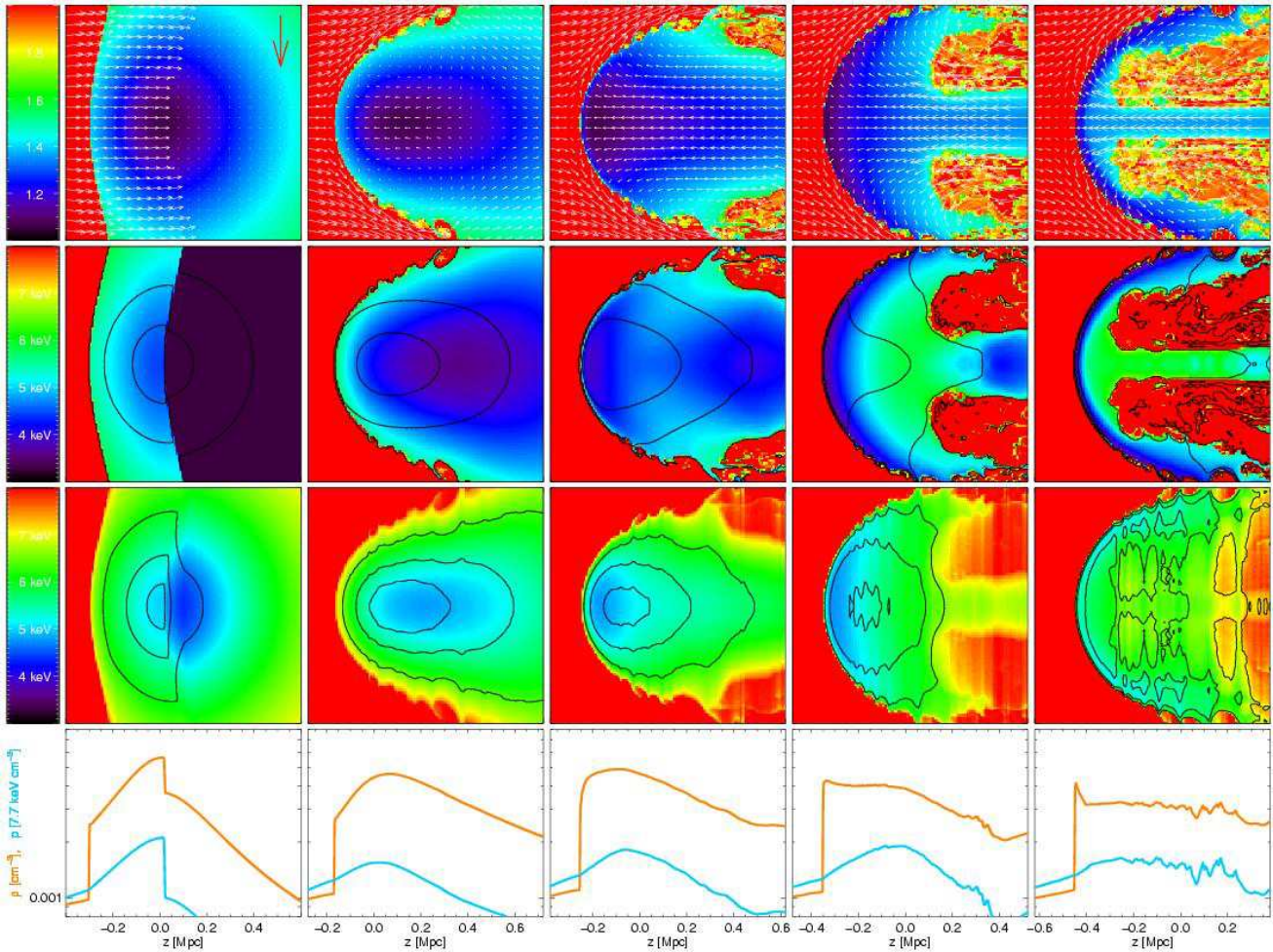}}
\caption{Simulation results for our fiducial run at time steps 0, 1, 2, 3,
and 4 Gyrs from left to right. Top panel: 2D slice of the entropy
$S=\log{\left[\left(p/\rho^{5/3}\right)/\left({\rm
keV\,cm}^{2}\right)\right]}$ (colour scale), overlayed with vector
representation of the velocity field (red arrow corresponds to a velocity
of 1400 $km\,s^{-1}$). Second panel: Slice through the temperature
distribution, overlayed with density contours. Third panel: Emission
weighted, projected temperature (assuming uniform background at relative
level and temperature measured by {\em Chandra}), overlayed with surface
brightness contours, assuming a side-on viewing angle of $90^{\circ}$.
Each box is 1 Mpc on a side. Bottom panel: plot of pressure (blue) and
density (red) along the axis.\label{fig:simulation}}
\end{figure*}

As material is streaming towards the front of the cloud, a developing
shear-layer due to unstable Kelvin-Helmholtz modes drives material along
the contact discontinuity backwards, in the direction of the background
flow.  Together with the backflow induced by the displacement of the cloud
core away from the center of the gravitational well, this results in the
formation of a strong vortex inside the cloud, which transports material
inside the cloud against the direction of the background flow along the
symmetry axis (i.e., to the left in Fig.~\ref{fig:simulation}), and, along
the surface, in the direction of the background flow (i.e., to the right in
Fig.~\ref{fig:simulation}).  It is this vortex that leads to the transport
of low entropy, low temperature material towards the front of the cloud.
We have visualised this transport in an animation viewable at {\small {\tt
http://www.mpa-garching.mpg.de/$\sim$heinzs/transport.html}}.
Morphologically very similar structures have been seen in previous
simulations of ram pressure stripping by \citet{balsara:94,murray:93}.

The effect of the internal circulation on the entropy and temperature
structure of the cloud can be seen in Fig.~1. The lowest entropy material
is transported to the front, while material with higher entropy fills the
center of the cloud.  The rarefaction wave behind the cloud leads to rather
low temperatures on the trailing side as well, due to the adiabatic
expansion of the material in the rarefaction.  However, the emission
weighted temperature map shows clearly that the dense, high emission
measure material at the front of the cloud produces the lowest observable
projected temperature.

\subsection{Transport processes and cloud destruction} 
In the outskirts of the cloud, the background flow can unbind gas from the
cloud potential as long as the ram pressure exceeds the binding energy of
the gas, i.e, as long as $\rho_{\rm x}\frac{v_{\rm x}^2}{2} \gtrsim
\frac{3}{5}\rho(r)\,c_{\rm s}^2$.  This sets the initial size $r_0$ of the
stripped cloud, after the passage of the wind driven shock wave (roughly
corresponding to the second panel in Fig.~\ref{fig:simulation}):
\begin{equation}
	r_{0} \sim r_{\rm c}\sqrt{\left[\left(\frac{c_{\rm s}}{v_{\rm
	x}}\right)^2\frac{2\rho_{\rm c}}{\gamma\rho_{\rm
	x}}\right]^{\frac{2}{3\beta}} - 1} \label{eq:radius}
\end{equation}
where $c_{\rm s}$ is the sound speed of the unperturbed, cold, isothermal
cloud, $\gamma=5/3$ is the adiabatic index of the gas, and $v$ is the
velocity of the external wind.  We define $\rho_0$ as the cloud density at
$r_0$.  In the case of the above simulation, $r_0 \sim 2 r_{\rm c}$.  For
large velocities or small density contrasts, all the material is unbound
and stripping is effective even in the absence of dynamical instabilities.
This is the case if $\rho_{\rm x}\frac{v^2}{2} \gtrsim \frac{3}{5}\rho_{\rm
c}\,c_{\rm s}^2.$ In the simulation shown above and, most likely in A3667,
this condition is not met \citep{vikhlinin:01b}.  Thus additional dynamical
effects must come into play to destroy the cloud.

The removal of material along the cloud surface is facilitated by
Kelvin-Helmholtz eddies \citep{nulsen:82}.  The largest growing mode has a
wavelength of order $\lambda_{\rm max} \sim r_{0}$ for $r_{0} > r_{\rm c}$.
In the limit of $\rho_{\rm x}\ll \rho_0$ (which is achieved both in
observed cold fronts and the simulation shown above), the growth time of a
mode of wavelength $\lambda$ is $\tau_{\lambda} =
\frac{\lambda}{v}\,\sqrt{\frac{\rho_{0}}{\rho_{\rm x}}}$. A shear layer of
thickness $\Delta \sim \lambda/2\pi$ set up by a fully non--linear mode of
wavelength $\lambda$ that transports material away from the cloud surface
with velocity of order $\lesssim v_{\rm x}$.  Since the largest unstable
mode has $\lambda_{\rm max} \sim r_0$, the destruction time of the cloud is
approximately given by
\begin{equation}
	\tau_{\rm dest} \sim {\rm few} \times \frac{2\pi r_0}{\lambda_{\rm
	max}}\frac{r_0}{v_{\rm x}} \sim {\rm few} \times 10\,
	\frac{r_0}{v_{\rm x}} \equiv \frac{r_0}{\zeta v_{\rm x}}
	\label{eq:destruction}
\end{equation}
where we defined the ablative index $\zeta$ to be calibrated by the
simulations.

The shear layer induces a vortex inside the cloud, which becomes stronger
over time as the width of the shear layer grows as longer and longer
wavelengths become fully non--linear.  By construction, the time it takes
for this vortex to transport the lowest entropy material from the core to
the front is $\tau_{\rm turnover}\sim \tau_{\rm dest}$, since this is the
time for the shear layer to remove most of the mass of the cloud, and thus
to turn over the entire cloud.  It is therefore natural to assume that
during the late stages of cloud evolution the coldest material will find
itself close to the front of the cloud.

Accordingly, from the simulation shown above, we see that the cloud is
destroyed at about the same time that the low entropy gas has been
transported to the front.  The numerical experiment yields $\zeta \sim
0.08$, which is the order magnitude expected from
eq.~(\ref{eq:destruction}).

\section{Illustrative case -- A3667}
To further illustrate the relevance of the mechanism, considered above, to
the actual structure of the observed cold fronts, we calculated the
temperature structure of A3667 using XMM-Newton data. The front in A3667
\citep{vikhlinin:01} is the clearest front detected by Chandra.

\begin{figure} 
\plotone{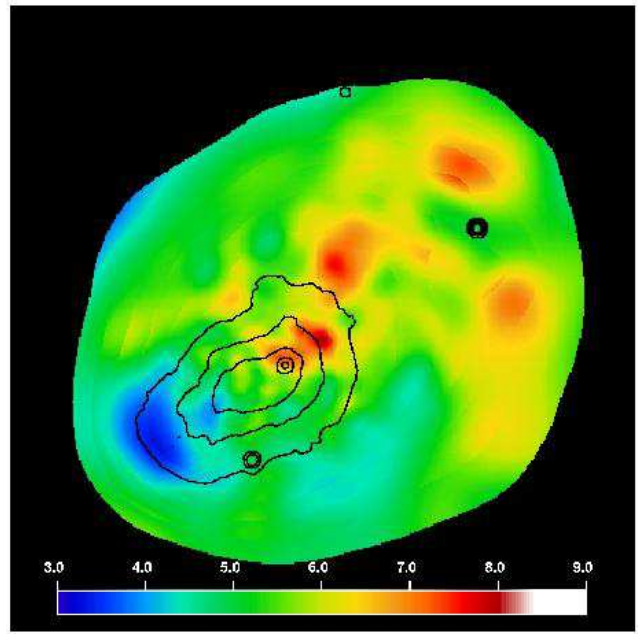}
\caption{Projected temperature distribution (in keV). The contours are the
isophotes of the X-ray surface brightness in the 0.5-3 keV energy band. The
outermost contour approximately traces the discontinuity, discussed in
\citet{vikhlinin:01}. The extension of the cool region outside the
discontinuity (outermost contour) is due to the significant smoothing of
the temperature map.  The image is 26 by 26 arcminutes.
\label{fig:xmm}}
\end{figure}

We used four XMM-Newton observations of A3667 (observations IDs 0105260301,
0105260401, 0105260501, 0105260601). The description of the data and
detailed analysis are given by \citet{briel:03}. Here we provide only the
gas temperature map for this cluster based on the MOS data. To calculate
the projected gas temperature distribution, we employed the method
described in \citet{churazov:99}. Namely, we fit the observed spectra at a
given location as a linear combination of two template spectra
corresponding to emission from an optically thin plasma (convolved with the
MOS energy response) with temperatures 2 and 8 keV and determine the gas
temperature as a function of the relative weights of the template
spectra. The resulting ``temperature map'', adaptively convolved to have an
effective number of counts per smoothing window of 6400, an approximate S/N
of 80, is shown in Fig.~\ref{fig:xmm}. The contours show isophotes of the
X-ray surface brightness in the 0.5-3 keV energy band. The outermost
contour approximately traces the discontinuity, discussed in
\cite{vikhlinin:01}. From this figure it is clear that the lowest entropy
gas is located along the interface and its temperature is lower than that
in any other place. The extension of the cool region outside the
discontinuity (marked by the contour) is due to the significant smoothing
used to generate the temperature map. The overall structure is strikingly
similar to the predictions of the simple model considered in the previous
section. It is therefore reasonable to assume that this cool gas is the low
entropy gas transported from the core of the cloud to the interface where
it expanded and cooled to the observed low temperature.

\section{Discussion and conclusion}
Cold fronts were also identified in the full (cold dark matter + fluid
dynamics) simulations of \cite{nagai:03} and \cite{bialek:02}. Compared to
these simulations our model is intentionally kept simple, using rather
idealised initial conditions. This simplicity, however, allows us to see
the evolution of the cloud characteristics in a physically more transparent
and controlled situation. The results shown above are in broad agreement
with the suggestions made by \cite{nagai:03} and \cite{bialek:02}, although
the detailed description of the cloud evolution is different.  As such, our
simulations are not meant to replace full blown 3D cluster merger
simulations, rather, we use them as a tool to point an important mechanism
at work inside ram pressure stripped sub-clusters, which, while present in
larger scale simulations, had not been identified in previous publications
on the subject.

We show that slow gas motion inside a gaseous subcluster moving through a
uniform medium transports gas from the central parts of the cloud towards
the contact discontinuity, which separates the subcluster gas from the
external gas. In a typical cluster or group of galaxies, the entropy of the
gas in the central region is significantly lower than the entropy of the
outer layers. The gas transported from the core region will undergo
adiabatic expansion and cool to temperatures less than its initial
temperature. This cool gas may form a cool and bright tip at the leading
edge of a moving cloud. Such a situation can be characteristic of a
subcluster or a group that still retains gas in its gravitational
potential, but spends sufficient time moving through the cluster gas so
that motions inside the cloud are fully developed.

Assuming that the original subcluster had a metallicity gradient with the
most metal rich gas near the center, the transport of the core gas towards
the interface should also enhance the abundance contrast across the
discontinuity.

As mentioned before, the simulations presented in \S\ref{sec:simulations}
ideallized for the purpose of isolating the process we were interested in
studying in this paper.  Many more levels of complexity can be added to
tailor the simulations to specific cluster merger scenarios.  For example,
it is easily possible to simulate such a merger by running two beta model
atmospheres into each other.  In this way, one can capture the additional
effects of varying ram pressure due to the stratified cluster atmosphere
and of tidal effects.

To verify that the effects discussed above act in such a more general
scenario, we ran several simulations of two beta model atmospheres
colliding, with one cluster being significantly larger, more massive, and
hotter than the other.  Indeed, the same mechanism discussed above operates
in such cases as well.  In addition, more complex dynamical effect occur in
the stripped wake, especially immediately after the passage of the smaller
cluster core through the bigger one.  This interaction also induces some
internal waves which produce interesting observable features inside both
atmospheres (comparable to the effects of sloshing mentioned above).
Additionally, adiabatic expansion of the material as it exits the larger
cluster after passage leads to a significant cooling of the material,
enhancing the effect described above even more: The lowest entropy and thus
coolest material lies at the front of the sub-cluster.

Such simulations introduce a number of additional free parameters, and it
is beyond the scope of this paper to map out all of this parameter space,
in particular since even such a simulation will be artificial in that it
neglects the evolution of the underlying dark matter, truncation effects of
the cluster atmospheres, and non-sphericity of the gas clouds.  We note,
furthermore, that reproducing a specific observational scenario in detail,
such as the example of Abell 3667, is a non-trivial exercise in fine
tuning.  For example, the estimated low Mach number of around $M\sim 1$
seen in this object \citep{vikhlinin:01} is difficult to reproduce by
simple free-fall simulations of two clusters, which ultimately lead to Mach
numbers of around $M\sim 3$ near the impact of the two cores, corresponding
to the typical depth of the potential well of a King model \citep{king:62}.
The frequency of incidence of such low Mach number merger shocks is a
question that can only be answered by detailed studies in the context of
structure formation simulations.

Thus, many other effects will play a role in determining the observational
appearance of subclusters in ram pressure dominated environments. For
example, a subcluster which has already passed through the dense core of
the main cluster will have been stripped of most of its outer gas.  Upon
leaving the core on the other side of the main cluster, the subcluster will
propagate into a lower pressure and lower density environment, thus
expanding adiabatically and cooling to form a cold cloud, as observed in
many cluster environments.  Mergers may also induce sloshing of low entropy
gas in the potential wells of the cluster cores and cause the displacement
of the cool gas relative to the equilibrium position
\citep{markevitch:01,markevitch:02,churazov:03}.  Nonetheless, the
processes described in this paper will operate under most of these
conditions as well, leading to the transport of cold material to the front
of the cloud on time scales comparable to the lifetime of the subcluster.

\thanks{We would like to thank Markus Brueggen, Torsten Ensslin, Maxim
Markevitch, Francesco Miniati, Alexey Vikhlinin, and Simon White for
helpful discussions. CJ and WF thank MPA for hospitality and support during
the summer of 2002 when this work was begun and also acknowledge support
from the Smithsonian Institution and NASA contract NAS8-39073.}

\label{lastpage}

\end{document}